 \documentstyle[12pt]{article}
\begin{document}
\parskip=6pt
\baselineskip=22pt
 \def\be{\begin{equation}}
\def\ee{\end{equation}}
\def\ba{\begin{array}}
\def\ea{\end{array}}
{\raggedleft{\sf${ ASITP}$-93-67\\}}
{\raggedleft{\sf{~~ revised.~${ June}$ 1994.\\}}}
\bigskip
\medskip
\centerline{\Large\bf The $\sigma$--Model and Non--commutative Geometry
\footnote{\sf Work supported in part by
The National Natural Science Foundation Of China.}}
\vspace{10mm}
\centerline{ \bf  Han-Ying Guo, ~~Jian-Ming Li~~and~~ Ke Wu  }
\vspace{1.5ex}
\centerline{\sf CCAST (World Laboratory), P. O. Box 8730, Beijing 100080,
China;}
\vspace{1ex}
\centerline{\sf  Institute of Theoretical Physics, Academia Sinica,
P. O. Box 2735, Beijing 100080, China.\footnote{\sf Mailing address}\\}
\vspace{4ex}

\centerline{\large \sf Abstract}

{\it In terms of non-commutative geometry, we show that the $\sigma$--model
 can be built up by the gauge theory  on discrete group $Z_2$. We introduce a
constraint  in the gauge theory,
which lead to the constraint imposed on linear $\sigma$ model to get nonlinear
$\sigma$ model .}

\section[toc_entry]{Introduction}
It has become widely accepted that strong interactions  exhibit a set of
approximate symmetries corresponding the chiral groups $SU_L(2)\times SU_R(2)$
and
$SU_L(3)\times SU_R(3)$. The best known Lagrangian model based on
$SU_L(2)\times SU_R(2)$ is the so called  $\sigma$-model[1]. There is a
quadruplet of real fields in the  linear $\sigma$ model. However,  the particle
corresponding one of them has not been observed experimentally. To meet this
difficulty, a constraint has to be imposed to eliminate this particle, then the
nonlinear $\sigma$ model was introduced. So far, nonlinear $\sigma$ models have
widespread applications in statistical mechanics as well as in quantum field
theory. Whether there  is  any profound meaning in
 $\sigma$-model from the ordinary geometrical point of view is an open
question.

Since Alain Connes[2] applied his non--commutative geometry to the particle
physics
model building in which the Higgs fields were introduced as gauge fields, many
efforts have been done alone  similar direction[3-6]. In particular, Sitarz[7]
developed discrete points idea and built a gauge theory  on discrete group.
 Soon after, the physical model building  alone this direction was completed by
the authors[8]. However,
the relationship between sigma model and non-commutative geometry has never
be explored.

In this letter, we show that the sigma  model may be regarded as a kind of
gauge field
in non-commutative geometry, based on
the Yang-Mills like gauge theory on discrete group developed by the present
authors[8].
The constraint imposed on the linear $\sigma$ model is introduced naturally by
the gauge theory, which may explore why nonlinear $\sigma$ model is the
realistic model.

\section[toc_entry] {Differential Calculus on Discrete Group}
In this section, we will outline the notion of differential calculus
theory on
discrete groups. For details, it is referred to [7].

Let G be a discrete group of size $N_G$ and ${\cal{A}}$  the algebra of
complex valued functions on G.
The right and left multiplications of G induce natural
automorphisms of ${\cal{A}}$. The right and left actions on ${\cal A}$ read
\be (R_{h}f)(g)=f(g\odot h),~
{}~~(L_{h}f)(g)=f(h\odot g), \ee
where $\odot$
denotes the group multiplication. The derivative on ${\cal A}$ is defined
as the left (or right ) invariant vector space  which
satisfies the following condition
\be
{\partial}\in {\cal F}\Leftrightarrow \{L_{h}\partial(f)=
\partial(L_{h}f), ~~\forall f\in {\cal A}\}.\ee
The basis of ${\cal F}, ~{\partial}_{i}, ~i=1, \cdots , N_{G} $ is given by
\be
{\partial}_{g}f=f-R_{g}f,~~g\in G',f\in {\cal A}, \ee
where  $G'=G \backslash e$, which is composed by the elements of group $G$
eliminated the unit $e$. As, ~${\partial}_{e}$, ~e the unit of G, is
trivial,
{}~~${\partial}_{e}f=0, ~~\forall f\in {\cal A},$
the dimension of ${\cal F}$ is then
$N_{G}-1$.
${\cal F}$ forms an algebra with relations
\be
{\partial}_{i} {\partial}_{j}={{\displaystyle\sum}_{k}}C_{ij}^{k}
{\partial}_{k}, \ee
where $C_{ij}^{k}$ are the structure constants satisfying
$${{\displaystyle\sum}_{l}}C_{ij}^{l}C_{lk}^{m}={{\displaystyle\sum}_{l}}
C_{il}^{m}C_{jk}^{l}$$
due to the associativity of the algebra ${\cal F}$.
It is easy to check the
following identity
$${\partial}_{g}{\partial}_{h}={\partial}_{g}+{\partial}_{h}-
{\partial}_{h\odot g}(1-\delta^e_{h\odot g} )~~~~ g,h \in G',$$
with structure constants
$ C^{g}_{gh}=1,~~~C^{h}_{gh}=1,~~C^{h\odot g}_{gh}=-(1-\delta^e_{h\odot g})$.
 In the case of $Z_{2}=\{e,r\}$, for example, the only nontrivial
structure constant is $C^{r}_{rr}=2$.

The Haar integral is introduced as a complex valued linear functional
on ${\cal{A}}$ that remains invariant under the action of $R_{g},$
\be {\int}_G f=\frac{1}{N_{G}}{\sum}_{g\in G}f(g),\label {g}\ee
which is  normalized such that ${\int}_G 1=1$.

\section[toc_entry] {Yang-Mills Like Gauge Theory on Discrete Group}
In this section, we introduce a free fermion  Lagrangian on $M\times Z_2$ and
build a Yang-Mills like gauge theory on $M\times Z_2$, in which  a simple
complex Higgs field
is the gauge field with respect to $ Z_2 $-gauge symmetry and the
Yukawa couplings between the fermions and Higgs  is obtained by
the
covariant derivatives or the minimum  gauge coupling.

To build the Yang-Mills like gauge theory, let us first assign the free fermion
field
with respect to $Z_2$
elements as follow
\be
\psi(x,e)=\sqrt{2}\psi_L(x),~~\psi(x,r)=\sqrt{2}\psi_R(x),~~e,r\in Z_2,
\ee
where
$${\psi}_{L}(x)=\frac{1}{2}(1+{\gamma}_{5}) \psi(x), ~~~{\psi}_{R}(x)=
\frac{1}{2}(1-{\gamma}_{5}) \psi(x)$$
are the left and right handed field respectively.

If we introduce a  Lagrangian on discrete group $Z_2$ as follow
\be
{\cal
L}(x,h)=\overline{\psi}(x,h)(i\gamma^\mu\partial_\mu+\mu\partial_r)\psi(x,h),
{}~~~~h\in Z_2
\label {a},\ee
we find that: $${\cal L}(x)=\int_{Z_2}{\cal
L}(x,h)=\overline{\psi}(x)(i\gamma^\mu\partial_\mu-\mu)\psi(x)$$
is  just the  Lagrangian of free fermion in space time $M^4$, then we call the
Lagrangian  in (\ref a) the free fermion Lagrangian on $M^4\times Z_2$.

Similar to the reason that leads to the introduction of Yang-Mills fields,
it is reasonable to require that the Lagrangian (\ref a) be invariant under
gauge
transformations $U(h), ~h\in Z_2$,
\be
\psi(x,h)\rightarrow {\psi(x,h)}'=U(h)\psi(x,h), \label {h} \ee
namely, we require $Z_2$ symmetry be gauged.
Generally  speaking, $U$ should be  functions on $M^4$ and discrete group
$Z_{2}$, to deal with the sigma model we confine our attention to the case that
they are  functions on discrete group.

Then for the second term
$\mu\overline{\psi(x,h)}{\partial}_{r}\psi(x,h)$ in (\ref a), it is
needed to introduce
gauge covariant derivative $D_{r}$ to replace ${\partial}_{r}$, which satisfy
\be
D_{r}\psi(x,h)\rightarrow
[D_{r}\psi(x,h)]'=U(h)D_{r}\psi(x,h)\label{2},\ee
in order that  $\overline{\psi}(x,h)D_{r}\psi(x,h)$ is
$Z_2$-gauge invariant. This can be realized if we introduce
a field $\phi(x,h)$, the Higgs field, as a connection with respect to
the $Z_2$-gauge symmetry and form the covariant derivative as follows
\be
D_{r}\psi=({\partial}_{r}+\frac{\lambda}{\mu} \phi R_{r}){\psi}.
\ee
Then the transformation law (\ref{2})
is satisfied if the generalized gauge field
$\phi(x,h)$ has the transformation property
\be
\frac{\mu}{\lambda}-\phi'=U(\frac{\mu}{\lambda}-\phi)(R_{r}U^{-1}).
\label{3}\ee

In ordinary Yang-mills gauge theory, we know that  we  can always choose a
gauge at any point
to make where the gauge
potential vanish. There is a theorem in the differential geometry:
For a connection in vector bundle, we  can choose a local frame at any point to
make the connection vanish at this point.

Accordingly, it is reasonable  to require  gauge field $\phi$ may be
transformed into zero at any point by choosing a special gauge. In view of
these consideration and
transformation rule (\ref  3), we obtain a constraint on the gauge field as
following:
\be
\phi=\frac{\mu}{\lambda}(1-U'^{-1}R_{r}U').
\label{31}\ee
where $U'$ is an  a  element of the gauge group.  We will find that this
constraint is useful in discussing the form of gauge field.

We introduce a new field $\Phi=\frac{\mu}{\lambda}-\phi$ such that the
transformation rule (\ref{3})  becomes
\be
{\Phi \rightarrow \Phi}'=U{\Phi}({R}_{r}U^{-1}).\label {m} \ee
and the new field $\Phi$ has the form as
\be
\Phi=\frac{\mu}{\lambda}U'^{-1}R_{r}U'.\label {32}\ee

 Similar to the usual gauge theory where the covariant derivative
is equivalent to the covariant  exterior derivative
$D_M=d_M+igA_{\mu}dx^{\mu}$
and $D_Mf=D_{\mu}fdx^{\mu}$, for the case in hand,
the covariant exterior derivative
takes form
\be
D_{Z_2}=d_{Z_{2}}+\frac{\lambda}{\mu}\phi{\chi}^{r}.\ee
 The reason is that
$$(d_{Z_2}+\frac{\lambda}{\mu}\phi {\chi}^{r})f=
({\partial}_{r}+\frac{\lambda}{\mu}
\phi R_{r})f{\chi}^{r}=D_{r}f{\chi}^{r}.$$

Thus from (\ref a), it follows the generalized gauge invariant Lagrangian for
fermions in each sector characterized by $Z_2$
\be
\begin{array}{cl}
{\cal{L}}_{F}(x,h)
&=\overline{\psi(x,h)}\{i{\gamma}^{\mu}{\partial}_{\mu}
+(\mu{\partial}_{r}+
\lambda \phi(x,h)R_{r})\}\psi(x,h) \\[4mm]
&=\overline{\psi}(x,h)(i\gamma^\mu{\partial_{\mu}}+\mu D_{r})\psi(x,h).
\end{array}\label{4}
\ee
The Hermitian property of operator $\phi R_{r}$ requires
that
\be
{\phi}^{\dag}(x,e)=R_{r}\phi(x,e)=\phi(x,r).\ee
After integrating the last terms in ${\cal L}(x,h)$ over $Z_{2}$ and in terms
of $\Phi$, we get
\be
{\int}_{Z_{2}}\mu\psi(x,h)D_r\psi(x,h)=
-\lambda\overline{\psi}_L (x)\Phi(x){\psi}_{R}(x)
-\lambda\overline{\psi}_R (x){\Phi}^{\dag}(x){\psi}_{L}(x)\ee
which is nothing but the Yukawa couplings between the Higgs and chiral
fermions.

{}From direct calculation, similar to Yang-Mills gauge theory,  $F_{r \mu}$ and
$F_{rr}$
are related to the covariant
derivatives respectively
\be
[D_{r},\partial_{\mu}]\psi=\frac{\lambda}{\mu} F_{r\mu}R_{r}\psi=
-\frac{\lambda}{\mu} F_{\mu r}
R_{r}\psi,\ee
$$(D_{r}D_{r}-2 D_{r})\psi=\frac{{\lambda}^{2}}{{\mu}^{2}} F_{rr}\psi. $$
where
$$F_{r \mu}={\partial}_{\mu}\Phi,~~~F_{rr}=\Phi{\Phi}^{_{\dag}}-
\frac{{\mu}^{2}}{{\lambda}^{2}}.$$
Under gauge transformation (\ref h), it is easy to see that
$$F_{r\mu}'=UF_{r\mu}(R_r U^{-1}),~~~
F_{rr}'=UF_{rr}U^{-1}.$$
Then we  can write down the gauge invariant  Lagrangian for the Higgs
field as following:
\be\begin{array}{cl}
{\cal L}_{H}(x,h)=
&TrF_{\mu r}{F}^{\mu\dag}_{r}
-\eta TrF_{rr}{{F}^{\dag}_{rr}}.
\end{array}\label{6}\ee
where $\eta$ is a positive real constant.
All the above identities can be  derived by non-commutative geometry[7,8]

Therefore, from (\ref 4 )and ( \ref 6), we get the entire Lagrangian of the
system
\be
{\cal L}(x)=\int_{Z_2}({\cal L}_{F}(x,h)+
{\cal L}_{H}(x,h))={\cal {L}}_{F}+{\cal {L}}_{H}
\label{8}\ee
where ${\cal {L}}_{F}$ is  gauge invariant part  for fermions with Yukawa
couplings to Higgs and ${\cal {L}}_{H}$ is the one for Higgs field
\be
\begin{array}{cl}
{\cal {L}}_{F}
=&\overline{\psi(x)}i{\gamma}^{\mu}{\partial}_{\mu}\psi(x)
-\lambda \overline{\psi}_{L}(x)\Phi(x) {\psi}_{R}(x)-\lambda \overline{\psi}_
{R}(x){\Phi(x)}^{\dag}{\psi}_{L}(x),\\[4mm]
{\cal {L}}_{H}=&
Tr(\partial_{\mu}\Phi(x))(\partial^{\mu}\Phi(x))^{\dag}-\eta Tr(
\Phi(x){\Phi(x)}^{\dag}-\frac{{\mu}^{2}}
{{\lambda}^{2}})^{2}.
\end{array}\label{9}\ee

In this work, the most important idea is that the Yukawa coupling is introduced
as gauge coupling. In doing so, Higgs field exists even in the absence of
Yang-Mills fields.
 As an example, we will study $\sigma$ model in next section.

\section[toc_entry]{$SU_L(2)\times SU_R(2)$ $\sigma$ model}
For the nucleon doublet field $N=\left(\ba{c}p\\n\ea\right)$, we set
\be
 \psi(x,e)=\sqrt{2}N_L,~~~\psi(x,r)=\sqrt{2}N_R. \ee
 and require  Lagrangian
$$
{\cal
L}(x,h)=\overline{\psi}(x,h)(i\gamma^\mu\partial_\mu+\mu\partial_r)\psi(x,h),
{}~~~~h\in Z_2 $$
is invariant under gauge transformations
\be
\psi(x,h)\rightarrow {\psi(x,h)}'=U(h)\psi(x,h),\label {j}\ee
where $U(e)\in SU_L(2)$ and $U(r)\in SU_R(2)$ are global
to the space time, but are local to the discrete group.

 It is clear that the elements of gauge group are independent  on the space
time but they are functions on the discrete group,  therefore, the gauge
invariant requires the
 scalar field $\Phi$ must
 exist. From equation (\ref 8), we have  the lagrangian for this model as
following:
\be\ba{cl}
{\cal L}=&i\overline{\psi}(x)\gamma^\mu\partial _\mu \psi(x)-\lambda
\overline{\psi}_L(x)\Phi(x)\psi_R(x)-\lambda \overline{\psi}_R(x)
\Phi^{\dag}(x)\psi_R(x)\\[4mm]
&+\partial_\mu \Phi(\partial^\mu\Phi)^{\dag}-\eta (\Phi\Phi^{\dag}-\frac
{\mu^2}{\lambda^2})^2\ea.\label {s}\ee

 From eq(\ref {32}), we obtain a constraint on the gauge field as following:
\be
\Phi=vSU_L(2)^{-1}SU_R(2)\ee
where $v=\frac{\mu}{\lambda}$.
 We know that elements of $SU(2) $ group may be written as
\be
G=\eta+i \tau\cdot \epsilon, \forall G\in SU(2)\ee
where $\tau_i, i=1, 2, 3$ are three Pauli matrices, $\eta$, $\epsilon$ are real
and
satisfy the unimodular condition $\eta^2+|\epsilon|^2=1$ .
Then we may have the   form of gauge field $\Phi$ as follow:
\be \Phi=v SU_L^{-1}(2)SU_R(2)=\hat\eta+i\tau\cdot \hat\epsilon \ee
where $\hat \eta$, $\hat \epsilon$ are real who
satisfy  condition $\hat\eta^2+|\hat\epsilon|^2=v^2$.

{}From above arguments, we know that the gauge field $\Phi$ valued on a sphere
surface of $U(2)$ group and the
radius of  the sphere is $v=\frac \mu\lambda$. Hence, we can write $\Phi$ as
$2\times 2$  matrix
\be
\Phi(x)=\sigma I+i\tau^i\pi_i\ee
where   $\sigma$ and ${ \pi}$ are real scalar fields  and satisfy a constraint
\be \sigma^2+|\pi|^2=v^2. \label {z}\ee It is obvious that this constraint is
remained under the gauge transformation. In following, we will show that if
 this constraint does not be taken into account, the linear  $\sigma$ model
may be obtained,
otherwise the nonlinear $\sigma$ model may be reached.

Rewrite the  Lagrangian (\ref s) in terms of the fields ${ \sigma, \pi}$, we
get
\be
{\cal L}=i\overline{N}\gamma^\mu\partial_\mu N-\lambda\overline{
N}(\sigma-i\gamma^5 {\tau\cdot \pi})N+
(\partial_\mu\sigma)^2+(\partial_\mu \pi)^2-\eta
(\sigma^2+\pi^2-\frac{\mu^2}{\lambda^2})^2.\label {y}\ee

The next step is to discuss the transformation property of the fields. If
desired one can introduce the  infinitesimal
gauge transformation as follows,
$$ U(h)=1+if(h)\cdot \tau, ~~f\in {\cal A},~~ h\in Z_2.$$
and set $f(e)=\alpha$, $f(r)=\beta$. Explicitly, we have
\be
U(e)=SU_L(2)=1+i\alpha\cdot \tau,~~ U(r)=SU_R(2)=1+i\beta\cdot \tau.\label
{x}\ee
 Under the gauge transformation  (\ref x), from (\ref j) and (\ref m) we have
\be\ba{cl}
&N\rightarrow N+i\frac{\alpha+\beta}{2}\cdot \tau
N+i\frac{\alpha-\beta}{2}\cdot \tau \gamma^5 N\\[4mm]
&\sigma+i\pi\cdot \tau\rightarrow  \sigma-(\alpha+\beta)\cdot \pi+i[\pi-\sigma
(\alpha+\beta)-(\alpha-\beta)\wedge \pi]\cdot \tau\ea.\ee
Redefine the parameter as $\alpha'=\alpha+\beta$,$\beta'=\beta-\alpha$, we have
\be\ba{cl}
&N\rightarrow N+i\frac{\alpha'}{2}\cdot \tau N-i\frac{\beta'}{2}\cdot \tau
\gamma^5 N\\[4mm]
&\sigma\rightarrow  \sigma-\alpha'\cdot \pi\\[4mm]
&\pi\rightarrow \pi-\sigma
\alpha'+\beta'\wedge \pi\ea\ee
which is just those for linear $\sigma$ model.

 In realistic model the constraint (\ref z) has to be required, we can use  it
 to eleminate the $\sigma$ field from the Lagrangin (\ref y), obtaining the
nonlinear $\sigma$ model,
\be
{\cal L}=i\overline{N}\gamma^\mu\partial_\mu N-\lambda\overline{
N}(\sqrt{v^2-\pi^2}-i\gamma^5 {\tau\cdot \pi})N+
(\partial_\mu \pi)^2+\displaystyle\frac {(\pi\cdot \partial
\pi)^2}{v^2-\pi^2}\ee

 In modern field theoy, the problem with the linear sigma model is the particle
corresponding to the field $\sigma$, which has not been observed
experimentally. One possibility
to get rid of this particle is to realize the chiral group $SU(2)$ nonlinearly,
by imposing the constraint (\ref z). However, as we have shown, once the
$\sigma$ model is introduced by means of non-commutative geometry, the
constraint is introduced naturally.  This may explore  why nonlinear $\sigma $
model is realistic model.

At last, we emphasize that as the Higgs fields, the $\sigma$ model has its
geometrical origin. Similarly,
other effective  Lagrangian models  can  be studied in terms of non-commutative
geometry also. We will discuss these topics in the future publications.

\bigskip

\centerline{\large \bf {Acknowledgements}}

\noindent  The author would like to thank H.B. Fei and Y. K. Lau
for helpful discussions.

\newpage

\centerline{ \bf {References}}
\bigskip

\begin{enumerate}

\item J.Schwinger, Ann.Phys.{\bf 2},(1958)407; \-M.Gell-Mann and M.Levy, Nuovo
Cimento\-
{\bf 16}(1960)705.

\item A. Connes, in: The Interface of  Mathematics and Particle Physics,\\
eds. D. Quillen, \- G. Segal and S. Tsou \- (Oxford U. P, Oxford 1990);
\-A. Connes and J. Lott, \- Nucl. Phys. (Proc. Suppl.) {\bf B18}, 44 (1990);
\-A. Connes and Lott, \- Proceedings of 1991 \-Cargese Summer \- Conference;
\- See also A. Connes, \-{\it Non-Commutative \- Geometry.}~~{\bf
IHES/M}/93/12.

\item D. Kastler, Marseille, CPT preprint {\bf CPT-91}/P.2610, {\bf
CPT-91}/P.2611.

\item R. Coquereaux, G. Esposito-Far\'ese and G Vaillant, Nucl Phys {\bf B353}
 689 (1991).

\item M. Dubois-Violette, \-  R. Kerner, J. Madore, J. Math. Phy.
{\bf 31}. (1990) 316;\-
B. S. Balakrishna, F G\"ursey and K. C. Wali, \-  Phys Lett {\bf B254}, 430
(1991).

\item A. H. Chamseddine, \- G Felder and J. Fr\"ohlich, \- Phys. Lett. \-
 {\bf 296B} (1993) 109.

\item A. Sitarz,  Non-commutative Geometry and  Gauge Theo\-ry on Dis\-crete
Groups, preprint {\bf TPJU}-7/1992;  A.Sitarz, Phys. Lett {\bf 308B}(1993) 311.

\item  Haogang Ding, Hanying Guo, Jianming Li and Ke Wu,
 Commun. Theor. Phys. {\bf
21} (1994) 85-94; J. Phys. A:Math.Gen.  {\bf 27}(1994) L75-L79;  J. Phys.
A:Math.Gen.  {\bf 27} (1994)L231-L236.

\end{enumerate}
\end{document}